\documentclass[aps,prb,twocolumn,showpacs,floatfix,superscriptaddress,amsmath,amssymb]{revtex4}
\usepackage{amssymb, amsmath}
\usepackage{verbatim}
\usepackage{graphicx}

\begin{document}
\title{Electronic interferometer capacitively coupled to a quantum dot}
\author{Seok-Chan Youn}
\affiliation{Department of Physics, Korea Advanced Institute of
Science and Technology, Daejeon 305-701, Korea}
\author{Hyun-Woo Lee}
\affiliation{PCTP and Department of Physics, Pohang University of Science and Technology, Pohang
790-784, Korea}
\author{H.-S. Sim}
\affiliation{Department of Physics, Korea Advanced Institute of
Science and Technology, Daejeon 305-701, Korea}
\date{\today}

\begin{abstract}
We theoretically study electron interference
in a ballistic electronic interferometer capacitively coupled to a quantum dot.
The visibility of the interference is reduced when the dot has degenerate
ground states with different excess charges.
The degree of the reduction depends on system parameters such as
the strength of the capacitive coupling,
and the dependence is analyzed
in the regime where the dwell time of electrons in the dot is much longer
than the electron flight time through the interferometry region
coupled to the dot.
The result is consistent with recent experimental data.
\end{abstract}



\maketitle

\section{Introduction}
Electronic interference is one of the main issues in mesoscopic physics.
It has been experimentally studied in various systems such
as a quantum ring\cite{Hansen01,Meier04} and an electronic Mach-Zehnder interferometer
(EMZI).\cite{Ji,Litvin,Roulleau}


The interference is reduced by the dephasing and by the phase averaging.
In the dephasing, an electron loses its phase due to the interactions
with other particles during its propagation along interference paths.
Electron-electron interaction is known to be the main source
of the dephasing at low temperature. The properties of the
interaction-induced dephasing vary from systems to systems.\cite{Seelig01, LeHur05, Kim08, Youn08, Neder08, Neder06b, Sukhorukov}
On the other hand, in the phase averaging, each electron does not
lose its phase but experiences a different phase shift. Then, the measured interference signal, which is the average over electrons, can be suppressed.
The phase averaging appears even in noninteracting systems under a finite bias voltage
or at finite temperature, as electrons in the finite energy window have different momentum and
thus different phase shifts.

Recently, Meier and the coworkers experimentally investigated
a quantum ring capacitively coupled to a
quantum dot in the Coulomb blockade regime.\cite{Meier04}
They measured the amplitude of the Aharonov-Bohm interference
of the ring, and simultaneously detected
electron current along a separate circuit containing the dot.
The amplitude shows a dip, when the dot has degenerate
states with different excess charges and thus shows a Coulomb-blockade peak of electron
conductance.
The dip was qualitatively explained~\cite{Meier04} as the phase averaging,
in which the phase accumulation of an electron along the ring depends on the occupation of the dot.
A quantitative analysis of the interference reduction, which has not been reported yet and is the aim
of the present work, will be
interesting and useful, as it comes from charge fluctuations in a {\em single} resonance level,
rather than those in macroscopic systems, thus contains the information
of the resonance.
Note that there have been a theoretical study~\cite{Seelig01}
on the dephasing
due to a macroscopic gate nearby the interferometer,
and
lots of studies\cite{QDdecoherence} on the properties
(such as decoherence) of the dot affected by its capacitive coupling to
a conducting wire.

\begin{figure} [tb]
\includegraphics[width=0.4\textwidth]{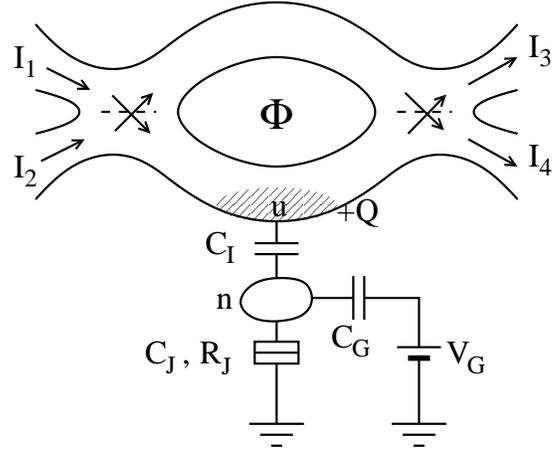}
\caption{\label{fig:1} Schematic diagram of an electronic interferometer
capacitively coupled to a quantum dot.
It couples to four reservoirs via two beam splitters (dashed lines)
and has two arms enclosing magnetic flux $\Phi$.
Here, multiple circulation paths along the interference loop are ignored;
thus the setup can be regarded as a Mach-Zehnder type.
The dot is in the Coulomb blockade regime, and capacitively couples to a (shady) region of the lower arm
(with capacitance $C_{\rm I}$)
and to a macroscopic gate (with $C_{\rm G}$).
Voltage $V_{\rm G}$ is applied to the gate.
The dot also connects to an electron reservoir
via tunneling junction with junction resistance $R_J$ and capacitance $C_{\rm J}$.
The fluctuation of charge $ne$ in the dot causes that of induced charge
$Q$ and potential $u$ in the coupling region of the lower arm, resulting in the
reduction of the interference visibility.}
\end{figure}

In this paper, we consider an electronic interferometer (Fig.~\ref{fig:1})
capacitively coupled to a quantum dot in the Coulomb blockade regime,
and study the reduction of
the interference visibility $\mathcal{V}$ due to the charge fluctuation of the dot;
other sources~\cite{Seelig01} of the reduction are omitted here.
We focus on the regime of $t_{\rm fl} \ll \tau_{\rm dwell}$, where
the reduction results from the phase averaging
rather than the dephasing.
Here, $\tau_{\rm dwell}$ and $t_{\rm fl}$ are
electron dwell time of
the dot and electron flight time $t_{\rm fl}$ through
the interferometer region coupled to the dot, respectively.

We derive the visibility $\mathcal{V}$ in the linear response regime, based on a self-consistent
treatment of the interaction, and
obtain the dependence of $\mathcal{V}$ on
temperature $k_B T$, gate voltage $V_\textrm{G}$
applied to the dot, and interaction strength parameter $g$
[defined below Eq.~\eqref{eq:ufreq}].
The visibility shows a dip when the dot shows a Coulomb-blockade resonance.
The depth of the dip is governed by $g$
while the width is determined mainly by $k_B T$.
The width is comparable with that of the Coulomb blockade
conductance peak
of the dot.
These results are in good agreement with the
experimental data.~\cite{Meier04}

%
%

Our setup is shown in Fig.~\ref{fig:1}.
Source-drain bias voltage $V_{\rm sd}$ is applied to reservoir 1.
Each arm of the interferometer is modeled by a disorder-free single-channel quantum wire
with length $L$ and the Fermi velocity $v_F$.
The length of the interferometer region coupled to the dot is denoted as $l$.
The dot is considered to be in the metallic Coulomb blockade regime where
its single-particle level spacing is much smaller than $k_B T$.
In this regime,
$E_C \gg k_B T, \hbar / \tau_\textrm{dwell}$ and
$R_{\rm J} \gg R_{0}$,
where $R_{0} \equiv h/e^2$ and $E_C = e^2 / C_\textrm{tot}$ is the charging energy of the dot [see Eq.~\eqref{eq:charging}].
We make the following simplifications.
We ignore multiple circulation paths around the interference loop, thus our model
can describe an EMZI.
In the regime of $\tau_\textrm{dwell} \gg t_\textrm{fl} (= l / v_F)$,
the potential $u$ of the coupling region
can be approximated~\cite{Seelig01} to be position-independent over the length $l$.
We ignore the effect of $C_{\rm J}$, since it is rather trivial, and will briefly discuss it later.


We remark that under the above simplifications, the RC time $\tau_\textrm{RC}$ in the coupling region
is much shorter than $\tau_\textrm{dwell}$, since $\tau_\textrm{RC} = R_0 C_\textrm{eff} < R_0 C_\textrm{tot}
= h / E_{\rm C}$ and $E_{\rm C} \gg \hbar / \tau_{\rm dwell}$,
where
$C_{\rm eff} \equiv C_{\rm G} C_{\rm I}/(C_{\rm G} + C_{\rm I})$ is the effective capacitance between the coupling
region and the gate.


\section{Potential fluctuation}

The fluctuation of the potential $u$ at the coupling region
affects the interference visibility.
Below, we obtain $u$,
by solving self-consistent
equations relating
$u$, the induced charge $Q$
in the coupling region, and the excess charge $- ne$ of the dot.
We note that $Q$ is well defined in our regime of $\tau_\textrm{RC} \ll \tau_\textrm{dwell}$
and that the same approach was used in Ref.~\onlinecite{Seelig01} where
an interferometer capacitively coupled to a macroscopic gate is considered.


In the frequency domain, $Q$ is related to
the dot charge $-n e$ and also to the bias voltage $V_{\rm sd}$,
$V_{\rm G}$,
\begin{eqnarray}
Q (\omega) & = & C_{\rm eff} [ u(\omega) -
V_{\rm G}(\omega) + \frac{e n(\omega)}{C_{\rm G}}], \label{eq:qrelat1}  \\
Q (\omega) & = & e^2 \nu (\omega) [V_{\rm sd}
(\omega)- N_{\rm res} u (\omega)]. \label{eq:qrelat2}
\end{eqnarray}
Here, $N_{\rm res}$ is the number of the reservoirs from which an electron can
move into the coupling region. It depends on the interferometer setup;
in the EMZI,~\cite{Ji} the channels
in the coupling region are chiral, thus, $N_{\rm res} = 2$, while
in a ring coupled to four reservoirs, the channels are not chiral and $N_{\rm res} = 4$.
The injectivity $\nu (\omega)$ is
the density of states of the electron entering from a reservoir to the coupling region
via the beam splitter in between.~\cite{Buttiker}
For simplicity, we have fixed the transmission probabilities of the beam splitters to be 0.5
so that the $N_{\rm res}$ reservoirs have the same injectivity,
$\nu (\omega) = i (1 - e^{i \omega t_{\rm fl}})/ (2 h \omega)$.
The origin of Eq.~\eqref{eq:qrelat2} is the fact that electron flow occurs
between the reservoirs and the coupling region, to screen excess
charges; the second term of Eq.~\eqref{eq:qrelat2} describes the screening.
From Eqs. \eqref{eq:qrelat1} and
\eqref{eq:qrelat2}, one finds
\begin{equation}
\label{eq:ufreq}
u (\omega) = \frac{(1-g^2) \nu (\omega) \frac{V_{\rm sd} (\omega)}{N_\textrm{res}}
+ g^2 \nu (0) [ V_{\rm G} (\omega) -  \frac{e n (\omega)}{C_{\rm G}}]}
{(1-g^2) \nu (\omega) + g^2 \nu (0)}.
\end{equation}
Here, $g^2 \equiv C_{\rm eff} /  (C_{\rm eff} + e^2 D)$ is
the dimensionless (Luttinger) parameter\cite{Blanter98} representing
interaction strength at the coupling region, and
$D = N_\textrm{res} \nu (0) = N_\textrm{res} t_{\rm fl}/2 h$ is the density
of states of the coupling region.


For $\tau_\textrm{RC}, t_\textrm{fl} \ll \tau_\textrm{dwell}$, 
Eq.~\eqref{eq:ufreq} predicts the following behavior of the time dependence $u(t)$.
When an electron tunnels into the dot so that $n$ changes from one integer $n_0$ to $n_0 + 1$,
$u(t)$ jumps from $u_{n_0}$ (value before the tunneling), oscillates around $u_{n_0 + 1}$,
and finally stabilizes to $u_{n_0 + 1}$. The oscillation decays roughly
\cite{footnote1} within the time scale of
$\tau_\textrm{RC}$. The stabilization
value $u_n = u(\omega = 0)$ is
\begin{equation}
u_{n} = (1-g^2) (V_{\rm sd}/N_\textrm{res}) +
 g^2 [V_{\rm G} - e n/C_\textrm{G}].
\label{eq:stabilized}
\end{equation}
Here, we ignore the fluctuation of the applied voltages $V_{\rm sd}$
and $V_{\rm G}$, which is valid\cite{Seelig01} in the case that the external circuit
connected to the interferometer has no impedance.
In the regime of $t_\textrm{fl} \ll \tau_\textrm{dwell}$, an interfering electron
in the setup feels the stabilization value $u_n$ of the potential most of time
so that the interference visibility is determined by the
discrete values $u_{n_0}$ and $u_{n_0+1}$ (see the next section).
In contrast,
as the system goes beyond the regime $t_\textrm{fl} \ll \tau_\textrm{dwell}$,
the visibility becomes affected by the variation of $u(t)$ between the stabilization values.



Using $Q (\omega)= C_\textrm{I} [u(\omega) - V_\textrm{dot}(\omega)]$,
one also finds
the electrostatic potential $V_\textrm{dot}$ of the dot as
$V_\textrm{dot} (\omega = 0) = [(C_\textrm{G} + g^2 C_\textrm{I}) (V_\textrm{G} - e n / C_\textrm{G})
+ C_\textrm{I} (1 - g^2) V_{\rm sd} / N_\textrm{res}]/(C_\textrm{G} + C_\textrm{I})$, when
the dot has $n$ excess electrons.
Then, the energy $E_\textrm{D} = \int^n_0 d n' (-e) V_\textrm{dot} (n')$ of the dot
is obtained as
\begin{equation}
E_\textrm{D} = \frac{(ne - Q_0)^2}{2 C_\textrm{tot}} + (\textrm{terms independent of $ne$}),
\label{eq:charging}
\end{equation}
the total capacitance of the dot is found to be $C_\textrm{tot} = C_\textrm{G} (C_\textrm{G} + C_\textrm{I}) / (C_\textrm{G} + g^2 C_\textrm{I})$,
and the effective gate charge is $Q_0 = C_\textrm{G} V_\textrm{G} +
(1-g^2) C_\textrm{I} C_\textrm{G} V_{\rm sd} / [N_\textrm{res} (C_\textrm{G} + g^2 C_\textrm{I})]$.
This result is understood from the fact that $u$ is affected
not only by the external voltages but also by screening.

It is worthwhile to discuss two limiting cases of $g \to 1$ and $g \to 0$.
The weak coupling limit of $g \to 1$ occurs when $C_\textrm{eff} \gg e^2 D$ (i.e., $\tau_\textrm{RC} \gg t_\textrm{fl}$).
In this limit, the charge fluctuation in the coupling region does not affect $u$, thus
$u \to V_\textrm{G} - en / C_\textrm{G}$ (independent of $V_{\rm sd}$),
$V_\textrm{dot} \to u$ (no charge accumulation in $C_\textrm{I}$),
and $C_\textrm{tot} (= e^2/E_\textrm{C}) \to C_\textrm{G}$.
On the other hand, in the strong coupling limit of $g \to 0$, $u$ is governed by $n$ and $V_{\rm sd}$,
and $C_\textrm{tot} \to C_\textrm{G} + C_\textrm{I}$.

For later use, we discuss the probability $P_n$ that
the dot has $n$ electrons.
In the Coulomb blockade regime, the dot states can be described by only
two occupation numbers, saying $n$ $\in \{ n_0, n_0+1 \}$, and have the energy
$E_\textrm{D} (n)$.
The probability $P_n$ is obtained\cite{Korotkov94,Utsumi}
from the stationary conditions, $P_{n_0} \Gamma^+ = P_{n_0+1} \Gamma^-$ and $P_{n_0} + P_{n_0 + 1} = 1$, as
\begin{equation}
P_{n_0} = \frac{1}{1+e^{\Delta_E /k_B T}}, \,\,\,\,
P_{n_0 + 1} = \frac{1}{1+e^{-\Delta_E /k_B T}}.
\label{eq:Pns}
\end{equation}
Here, $\Gamma^+$ ($\Gamma^-$) is the rate of the transition $n_0 \to n_0 + 1$ ($n_0 + 1 \to n_0$),
$\Gamma^{\pm} = \pm \Delta_E / [e^2 R_{\rm J} (1-e^{\mp \Delta_E / k_B T})]$,
and $\Delta_E = E_\textrm{D} (n_0) - E_\textrm{D} (n_0 + 1)
= e^2 (Q_0/e - n_0 - 1/2)/ C_\textrm{tot}$ is the energy difference between the two states
of $n_0$ and $n_0 + 1$. 
The dwell time $\tau_\textrm{dwell}$ of the dot can be estimated by
$\hbar / \Gamma^+$ for the state with $n_0$ ($\hbar / \Gamma^-$ for $n_0+1$).

\section{slow fluctuation regime}

In this section, the visibility of the interference in the current through the
interferometer is derived and analyzed in the regime of $t_\textrm{fl} \ll \tau_\textrm{dwell}$.

In the linear response regime, electron current from the source reservoir $1$ to reservoir 3 (one of the drains) can be described by the Landauer-B\"uttiker formula.
The differential conductance is given by
$G_{13} = (e^2/h) \int dE \left( - \partial f / \partial E \right) \overline{\left< T_{13} (E) \right>}$.
Here, $T_{13} (E)$ is the transmission probability of an electron with energy $E$ from the reservoir 1 to 3, and
$\overline{\left< \cdots \right>}$ denotes the statistical average over the fluctuation of $u$.

In the regime of $t_\textrm{fl} \ll \tau_\textrm{dwell}$, the occupation number of the dot
does not fluctuate most of time while an interfering electron passes through the coupling region.
In this case, and when the dot has $n$ excess electrons, $u$ has the stabilization value $u_n$ in
Eq.~\eqref{eq:stabilized}, and the resulting transmission probability $T_{13;n}$ has the form of
\begin{equation}
T_{13; n}(E) = \frac{1}{2}
\left\{ 1 + \cos \left[- 2\pi \Phi/\Phi_0 + \Theta_n \right] \right\},
\end{equation}
where $2\pi \Phi/\Phi_0$ is the Aharonov-Bohm phase, $\Phi_0 = h/e$,
and
$\Theta_n = -e u_n t_\textrm{fl} / \hbar$ is the phase shift induced by the potential $u_n$;
here, the lengths of the two arms
of the setup are assumed to be identical so that the dynamical phase is absent in $T_{13;n}$.
At the gate voltage where the probability $P_n$ is
not negligible only for $n \in \{ n_0, n_0 + 1 \}$ [see Eq.~\eqref{eq:Pns}],
one obtains
the ensemble average of $T_{13;n}$ over the possible values of $n$,
$\overline{\left< T_{13} (E) \right>} = \sum_{n = n_0, n_0+1} P_n T_{13;n}(E)$, and
the visibility $\mathcal{V} \equiv [\textrm{max}(G_{13}) - \textrm{min}(G_{13})]/[\textrm{max}(G_{13}) + \textrm{min}(G_{13})]$ of the interference in $G_{13}$,
\begin{eqnarray}
\mathcal{V} & = &  \lvert \sum_{n} P_n e^{-i \Theta_n} \rvert = \sqrt{1- \frac{2 \Gamma^+ \Gamma^- (1 - \cos \Delta \phi)}{(\Gamma^+ + \Gamma^-)^2}
}, \label{eq:vis} \\
\Delta \phi & \equiv & \Theta_{n_0 + 1} - \Theta_{n_0} =
\frac{4 \pi (1-g^2)}{N_\textrm{res}}
\frac{C_\textrm{I}}{C_\textrm{G} + C_\textrm{I}}, \label{eq:phase}
\end{eqnarray}
where $\textrm{max}(G_{13})$ ($\textrm{min}(G_{13})$) is the maximum (minimum) value of $G_{13}$ in one Aharonov-Bohm period
and $\Delta \phi$ is the phase difference between the two cases of $n_0$ and $n_0 + 1$.
Hereafter, we put $N_\textrm{res} = 2$, by considering an EMZI.~\cite{Ji}


\begin{figure}[tb]
\includegraphics[width=0.45\textwidth,height=0.22\textheight]{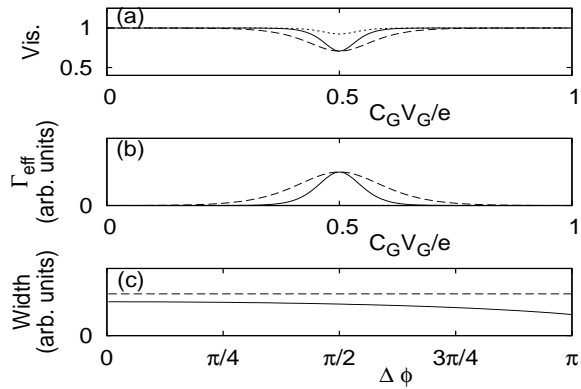}
\caption{\label{fig:2} (a) Visibility $\mathcal{V}$ and (b) effective
transition rate $\Gamma$ of the dot, as a function of $V_\textrm{G}$,
in the regime of $t_\textrm{fl} \ll \tau_\textrm{dwell}$.
$V_\textrm{G}$ varies around the Coulomb
blockade resonance where the dot has $n_0 = 0$ or $(n_0 + 1) = 1$ electrons.
Different values of temperature $t=k_B T/E_{\rm c}$ and phase difference $\Delta \phi$
are considered:
$(t, \Delta \phi)= (0.05, \pi/2)$ [see full line], $(0.1, \pi/2)$ [dashed], and
$(0.05, \pi/4)$ [dotted]. We put $N_\textrm{res}=2$.
(c) The width of the dip of $\mathcal{V}$ in (a) as a function of $\Delta \phi$
[see full line]. It is found to weakly depend on $\Delta \phi$.
For comparison, we also draw the width of the peak of $\Gamma_\textrm{eff}$ in (b) [dashed]; it is independent of $\Delta \phi$.
The sizes of the two widths are comparable.
}
\end{figure}


In Fig.~\ref{fig:2}(a), we plot the visibility $\mathcal{V}$ as a function of $V_\textrm{G}$.
Around the Coulomb blockade resonance of $\Delta_E = 0$, where the occupation number of
the dot maximally fluctuates ($P_{n_0} = P_{n_0 + 1} = 0.5$), $\mathcal{V}(V_\textrm{G})$ shows a dip.
The dip center occurs
at the gate voltage of $Q_0 = e(n_0 + 1/2)$.
The depth $d$ and the width $V_\textrm{width}$ of the dip are
\begin{eqnarray}
d = 1 - |\cos \frac{\Delta \phi}{2}|, \,\,\,\,\,\, e V_\textrm{Width} = 2 \xi k_{\rm B} T,
\label{eq:details}
\end{eqnarray}
where $\xi = \log (1+\eta)/(1-\eta)$, $\eta = \sqrt{3/4
- (1-\cos (\Delta \phi/2))/(1-\cos \Delta \phi)}$;
note that
$\mathcal{V} =
|\cos (\Delta \phi/2)|$ at the dip center.
The depth is independent of $k_B T$, and becomes larger for stronger interaction.
The dip (i.e., the reduction of the visibility) disappears in the weak coupling limit of $g \to 1$,
while $d \to 1 - \cos [(2\pi / N_\textrm{res}) C_\textrm{I} / (C_\textrm{G} + C_\textrm{I})]$ in the strong coupling limit of $g \to 0$.
The width $V_\textrm{width}$ linearly increases with the increase of the temperature,
and depends weakly on $g$;
as $\Delta \phi$ changes from $0$ to $\pi$, $\xi$ varies from $1.76$
to $1.10$.

On the other hand, the linear-response conductance of a circuit including the dot is known~\cite{Korotkov94,Utsumi}
to be proportional
to the effective transition rate
$\Gamma_\textrm{eff} \equiv \Gamma^+ \Gamma^- / (\Gamma^+ + \Gamma^-)$.
As a function of the gate voltage, it shows a peak at the Coulomb blockade resonance.
The width of the peak is proportional to the temperature; the full width
at half maximum is given by $4.35 k_{\rm B} T/e$.
Interestingly, the width of the conductance peak is comparable to that of the visibility dip
[Fig.~\ref{fig:2}(c)].
This finding is in good agreement with the experimental
result\cite{Meier04} of the visibility dips and the conductance peaks.

The mechanism of the above visibility reduction is the phase averaging.
In the regime of $t_\textrm{fl} \ll \tau_\textrm{dwell}$, each interfering electron does not lose
its phase while it passes along the coupling region, but its phase is either $\Theta_{n_0}$ or
$\Theta_{n_0 + 1}$, depending on the occupation of the dot.
The average over the different phases
cause the reduction of the interference signal.
This behavior is in contrast to the similar case~\cite{Seelig01}
of an interferometer coupled a macroscopic gate.
In this case, the reduction mechanism is the dephasing
due to thermal charge fluctuations, and
the details such as the temperature dependence of the reduction are different
from our result.

We note that the qualitatively same result is obtained for a semiconductor
dot where only one single-particle level is relevant.
And, multiple circulation paths along the interference loop, ignored in our work, will not
modify the above result qualitatively, as far as the time scales of the multiple circulations
of an electron
along the loop
are not much longer than $\tau_\textrm{dwell}$. This guarantees the agreement between
our result for the EMZI and the experimental data\cite{Meier04} for a ring.

In addition, our result is only qualitatively modified
in a more realistic case with finite $C_{\rm J}$ and a direct capacitance $C_{\rm D}$
between the coupling region and the gate.
In the presence of $C_{\rm J}$ and $C_{\rm D}$, Eq.~\eqref{eq:phase} is modified by the replacements,
$g^2 \equiv C_\textrm{eff}/(C_\textrm{eff} + e^2 D) \to C'_\textrm{eff}/(C'_\textrm{eff}+e^2 D)$,
$C_\textrm{eff} \equiv C_{\rm G} C_{\rm I} / (C_{\rm G} + C_{\rm I}) \to C'_\textrm{eff} \equiv C_{\rm D} + C_{\rm G} (C_{\rm I} + C_{\rm J})/(C_{\rm G} + C_{\rm I} + C_{\rm J})$,
and $C_{\rm G} + C_{\rm I} \to C_{\rm G} + C_{\rm I} + C_{\rm J}$ in the denominator.
The modifications in other capacitance-dependent quantities of $Q_0$ and $C_{\rm tot}$
(not shown here)
does not affect our main results in Eqs.~\eqref{eq:phase}-\eqref{eq:details} and in Fig.~\ref{fig:2}.

%
%

Finally, we briefly discuss the case that
the system deviates from the regime of $t_\textrm{fl} \ll \tau_\textrm{dwell}$;
this case is the strong coupling regime of $g \ll 1$
(i.e. $ t_\textrm{fl} \gg \tau_\textrm{RC}$) under the Coulomb blockade condition
of $\tau_\textrm{dwell} \gg \tau_\textrm{RC}$.
In this case, one has to take into account of the fluctuation of $u(t)$
over the flight time $t_\textrm{fl}$, thus our approach should be modified.
For example, for $\tau_\textrm{dwell} \gtrsim t_\textrm{fl}$,
the visibility and the dephasing rate may be approximately obtained,
as in Ref.~\onlinecite{Seelig01},
by studying the fluctuation spectra of $u$ and $n$,
and they will deviate from Eqs.~\eqref{eq:vis}-\eqref{eq:details}
due to the additional dephasing induced by the dependence of $u$ on time.
Note that the fluctuation spectra will be affected by the back-action from $u(t)$ to $n(t)$
except for $C_{\rm I} \ll C_{\rm G}$.
On the other hand,
the limiting case of $t_\textrm{fl} \gg \tau_\textrm{dwell}$
is likely to be hardly found in a realistic situation such as in Ref.~\onlinecite{Meier04}.
For this case, our assumption that $u$ has
no spatial dependence over the coupling region is not valid,
and
other approaches~\cite{Blanter98} may be helpful.

\section{conclusions}

We have investigated the reduction of the interference visibility in an electronic interferometer
capacitively coupled to a quantum dot. The reduction appears when the dot shows
Coulomb blockade resonances. The features of the resonant reduction of the visibility are different
from the dephasing effect in an interferometer coupled to a macroscopic gate.
Our result of the visibility reduction can be used for the detection of
single charges, and provide a
model for a local charge trap\cite{Litvin,Litvin08b}
(unintentionally) formed nearby an electronic interferometer.


We thank M. B\"{u}ttiker for useful discussion, and
KRF (2006-331-C00118), NRF (2009-0078437), and MOST (the leading
basic S \& T research projects) for support.


\bibliographystyle{apsrev}

\begin{thebibliography}{99}
\bibitem{footnote1} In the strong coupling regime ($g \ll 1$), the decay takes longer time
than $\tau_\textrm{RC}$ in our model. It may be the artifact of the simplification that the spatial
dependence of $u$ is ignored.
Even in this case, the phase shift $\int^{t_\textrm{fl}}
dt' u(t')$ stabilizes within $\tau_\textrm{RC}$, and thus the effect of the longer decay
on the visibility is negligible in the case of $t_\textrm{fl} \ll \tau_\textrm{dwell}$.
\end{thebibliography}

\end{document}